\documentclass[12pt]{article}
\usepackage{epsfig}

\begin{document}

\title {The birth of spacetime atoms as the passage of time\footnote{Article based on an invited talk at the conference ``Do we need a physics of passage?'' Cape Town, South Africa, 10-14 Dec 2012. 
 Submitted for publication in Annals of the New York Academy of Sciences (2014).
 }}

 \author{Fay Dowker\\ Blackett Laboratory, Imperial College, \\Prince Consort Road, London SW7 2AZ, UK}
  \maketitle
\begin{abstract}

The view that the passage of time is physical finds expression in the classical sequential growth models 
of Rideout and Sorkin in which a discrete spacetime grows by the partially ordered
accretion of new spacetime atoms. 
  
\end{abstract}
\vskip 1cm
 

\noindent\textbf{Introduction}

To answer the question ``Do we need a physics of passage?''  requires one to decide if there are relevant observational facts 
 unexplained by current science. There is no consensus on this. For example, there is a long-standing and persistent 
disagreement between those  -- let us call them ``Blockheads'' after J. Earman, as cited by J. Butterfield in this volume --  who
claim that our sense-experience of the passage of time can be adequately 
accounted for within a theory in which the physical world is a spacetime Block, 
and those who contend that there is \textit{something essential} about our sense-experience which corresponds
to nothing to be found within such a Block Universe picture \cite{Norton:2010}. 

One reason for this lack of agreement is that the observable ``fact''  in question, the experience
appealed to by those who are sceptical of the Block, is difficult to pin down scientifically. 
This difficulty can be partly explained if one accepts that we are
currently in a period of development of a new theory of spacetime without which 
the experience of passage cannot be properly communicated. For, the way we understand 
the physical world is through 
our best scientific theories and observational facts are most clearly communicated
if they are statements about concepts within an agreed theory of the world. 
There are no ``raw data" in science floating free of any theoretical framework. Science progresses when 
data accumulate that cannot be accommodated within the current best theoretical framework and the
framework itself must be re-made on new foundations. It is in the very nature of 
science, then, that data may look problematic and fragmentary in a period during which a 
new theory is being developed, before the framework exists within which they can be given proper expression.
It seems inevitable that during the period of discovery of a new theory, especially one that is 
profoundly revolutionary, there will be confusion, mis-communication and disagreement 
about what the observational ``facts'' actually are. 

Thus, I wish to suggest, our inability to reach consensus on the passage of time could be a consequence of
our not yet having made the necessary scientific  progress. 
We do not have a successful theory of spacetime that coherently incorporates the quantum nature of the
physical world, so we do not yet know the nature of the 
deep structure of spacetime. Some of the observational facts on which the new
theory will be built may, therefore, now be only roughly communicable and our sense-experience of the
passage of time may be an example of such a fact. 

In the last decades, however, progress on one approach to
finding a theory of quantum spacetime -- or quantum gravity as it is usually called -- 
affords us a forward look at how the passage of time may eventually
find a place in science. The approach, \textit{causal set theory},  is based on the hypothesis that 
spacetime is fundamentally granular or atomic at the Planck scale and this atomicity opens the door to new dynamical 
possibilities for spacetime and, hence, to a new perspective on the dichotomy of Being and Becoming.
In this article I will describe this progress and will 
expand upon R. D. Sorkin's claim that it gives us scientific purchase on the concept of passage
 \cite{Sorkin:2007qh}. 
 
\vskip .5cm   \noindent\textbf{Continuum spacetime is order plus scale}

In order to set the stage for the causal set hypothesis for the deep structure of spacetime, I will 
briefly survey our current best understanding of spacetime given to us by General Relativity 
 (see also C. Rovelli's contribution to this
volume).  Spacetime is a continuous, smooth, four-dimensional material that bends, warps and ripples according to
dynamical law as specified by the Einstein equations. Even when there is no matter present, ``empty'' spacetime 
itself can carry energy in the form of gravitational waves. Indeed this is the explanation for the 
variation in the rotation rate of the Hulse-Taylor binary pulsar system, which can be accurately modelled as a system 
losing energy via this gravitational radiation. The spacetime material is, however, 
very different from  those substances that populated pre-relativistic physical theory in that it is
intrinsically four-dimensional. It cannot be understood as a three-dimensional entity -- ``space'' -- 
evolving in time because that would imply a global time in which the evolution 
occurs and there is no such global, physical time in General Relativity (GR). 
 The notion that at one moment of time there is space, a 3-d geometry, and at the next 
moment space has evolved to another 3-d geometry is wrong in GR. 
There is no such physically meaningful entity as  3-d space, no physically meaningful 
slicing of spacetime into space-changing-in-time.

Having focussed on what spacetime in GR is \textit{not}, we can ask what it \textit{is}.
The structure of spacetime that takes centre stage in understanding the physics of GR is its 
\textit{causal structure}. This causal structure is a \textit{partial order} on the 
points of spacetime.\footnote{It is common to refer to the points of spacetime as \textit{events}.
I will not do so here because I will want to use event to mean something more general, something that may 
be extended in spacetime. A point of spacetime is an idealisation of an event that has 
no physical extent, like an infinitely small, infinitely fast-burning firecracker \cite{Geroch:1978}.} 
Given two points of spacetime, 
call them $A$ and $B$,  they will either be ordered or unordered.  
If they are ordered then one, let's say $A$ without loss of generality, 
\textit{precedes} -- or, is to the past of -- the other, $B$. This ordering is 
\textit{transitive}: if $A$ precedes $B$ and $B$ precedes $C$ then 
$A$ precedes $C$. The order is \textit{partial} because 
there are pairs of spacetime points such that
 there is no physical sense in which one of the pair precedes the other, they are simply unordered. This 
lack of ordering does \textit{not} mean the points of the pair are 
simultaneous because that would imply they occur at the same ``time'' and require
the existence of a global time for them to be simultaneous in.  Again: global time
does not exist in GR. 

This partial ordering of the points of spacetime is referred to as
the causal structure of spacetime because it coincides with the 
potential for physical effects to propagate. Physical effects can propagate
from $A$ to $B$ in spacetime if and only if $A$ precedes $B$ in the causal structure. 
If two spacetime points  are unordered then no physical effect can propagate from
one to the other because to do so would require something physical to travel faster than light. 
Causal structure plays a central role in GR and indeed the 
epitome of the theory, a black hole, is \textit{defined} in terms of the causal 
structure: it is a region of spacetime such that nothing that happens 
in that region can affect anything outside the region. 
 It is only by thinking of a black hole
in terms of causal structure that its physics can be understood.\footnote{As an 
example of this, it is very difficult to answer the question, ``Does someone falling feet first into 
a black hole lose sight of their feet as their feet cross the horizon?'' without 
drawing the conformal ``Penrose'' diagram that depicts the causal structure.} 

If there is no global, universal time, where do we find within GR the concept of 
physical time at all? Physical time in GR is associated, locally and ``personally,''
with every localised physical system, in a manner that more closely reflects our intimate 
experience of time than the global time of pre-relativistic Newtonian mechanics. 
 Each person or object traces out a trajectory or
\textit{worldline} through spacetime, a continous line in spacetime that is totally ordered by
the causal order: for any 2 points on the worldline, one precedes the other.
GR also provides a quantitatively precise concept of \textit{proper time} that elapses along each timelike
 worldline. A clock carried by 
a person following a worldline through spacetime will measure this proper time as it elapses, 
locally along the worldline. Viewed from this perspective, the famous ``twin paradox'' is no longer a paradox: two people who 
meet once, then follow different worldlines in spacetime and meet a second time in the future will in general
have experienced different amounts of proper time -- real, physical time -- elapsing along their different
worldlines between the meetings. Clocks are ``odometers for time'' along worldlines through spacetime. 

The remarkable thing, from this perspective, is that we get by in everyday life quite well under the 
assumption that there \textit{is} a global Now, a universal global time, 
and that we can synchronise our watches when we meet,
 then do different things and when we meet again our watches will still be synchronised. GR explains this
because it predicts that as long as the radius of curvature of spacetime is large compared to the physical
scale of the system and the relative velocities of the subsystems involved are small compared with the speed of light, 
the differences in proper time that elapse along our different worldlines will be negligible. 
We can behave \textit{as if} there is a global time, a global moment of Now, 
because for practical everyday purposes our clocks will remain synchronised 
to very high precision. 

In addition to being the key to understanding GR, the causal structure of spacetime is a unifying 
concept. Theorems by Kronheimer and Penrose \cite{Kronheimer:1967}, Hawking \cite{Hawking:1976fe} and Malament 
\cite{Malament:1977}
establish that  the causal order unifies within itself topology, including dimension, differentiable structure 
(smoothness) and almost all the spacetime geometry.  The only geometrical information
that the causal structure lacks is local physical scale.\footnote{Technically, the result states 
that if two distinguishing spacetimes are
causally isomorphic then they are conformally isometric. In 4 dimensions this implies
that the causal structure provides 9/10 of the metrical information as the metric is given by a symmetric 
$4\times 4$ matrix field of
10 spacetime functions, 9 of which can be fixed in terms of the 10th.} This local scale information can be furnished 
by providing the spacetime volume of every region of spacetime or, alternatively, the amount of proper time that 
elapses -- the \textit{duration} -- along every timelike worldline. In the continuum, the causal structure
and local  scale information complement each other to give the full geometry of spacetime, the complete 
spacetime fabric.  

\vskip .5cm   \noindent\textbf{2. Atomicity unifies order and geometry}

There are strong, physical arguments that the smooth manifold structure of spacetime must 
break down at the Planck scale where quantum fluctuations in the structure of 
spacetime cannot be ignored. The most convincing evidence that spacetime cannot 
remain smooth and continuous at arbitrarily small scales and that the scale at which the 
continuum breaks down is the Planck scale is
the finite value of the entropy of a Black Hole \cite{Sorkin:1985bu}.  Fundamental spacetime discreteness 
is a simple proposal that realises the widely held expectation that there must be a 
physical, Planck scale cutoff in nature. According to this 
proposal, spacetime is comprised of discrete ``spacetime atoms''  at the Planck scale.  

The causal set 
programme for quantum gravity \cite{Bombelli:1987aa, Sorkin:1990bj, Sorkin:2003bx} is based on the observation that
such atomicity is exactly what is needed 
in order to  conceive of spacetime as pure causal order since in a discrete spacetime,
physical scale -- missing in the continuum -- can be provided by \textit{counting}.  
For example, a worldline in a discrete spacetime would 
consist of a chain of ordered spacetime atoms and its proper time duration, in fundamental 
Planckian scale units of time of roughly $10^{-43}$ seconds, would be simply the \textit{number} of spacetime atoms 
that comprise the worldline. 

Causal set theory thus postulates that underlying our seemingly 
smooth continuous spacetime there is an atomic spacetime taking the form of a discrete, partially ordered
set or \textit{causal set}, whose elements are the spacetime atoms. 
The order relation on the set gives rise to the spacetime causal order in the 
approximating continuum spacetime and the number of causal set elements comprising a region 
gives the spacetime volume 
of that region in fundamental units. The Planckian scale of the atomicity means that there would be roughly $10^{240}$ 
spacetime atoms comprising our observable universe. 

According to causal set theory, spacetime is a material comprised of spacetime atoms which are, 
in themselves, featureless, with no internal structure and are therefore identical. Each atom acquires 
individuality as an element of a discrete spacetime, a causal set,   in view of its 
order relations with the other elements of the set. 
Let me stress here a crucial point:  the elements of the causal set, the discrete spacetime,
 are atoms of 4-d spacetime, \textit{not} atoms of 3-d space. An atomic theory of space would run counter to the
 physics of  GR in which 3-d space is not a physically meaningful concept. An atom of spacetime is 
 an idealisation of a click of the fingers, an explosion of a firecracker,  a here-and-now. 
 
 \vskip .5cm   \noindent\textbf{Growing discrete spacetime}

The previous paragraphs provided a sketch of the kinematics of causal set theory: spacetime is
a discrete order. The programme of causal set quantum gravity must now  provide the theory with a 
\textit{dynamics}, the analogue of the Einstein equations for spacetime in GR. In particular 
the programme faces the challenge of explaining  why the causal set that corresponds to the spacetime we observe around 
us is so special amongst causal sets. Most causal sets do not have nice smooth continuous spacetimes as approximations 
to them and the dynamics of causal set theory has to do the job of predicting that such non-continuum-like
causal sets do not occur. This is work in progress and current research is focussed on the path integral approach 
to quantum theory as the framework for this dynamics. Indeed, Dirac's choice of canonical quantisation 
or the path integral  \cite{Dirac:1933} for quantum mechanics is Hobson's choice for causal sets because the canonical 
approach requires a Hamiltonian and thus a continuous time variable, which causal sets manifestly do not provide: quantum causal 
sets demand the   path integral.
Path integral quantum theory is a species of generalised measure theory in which classical stochasticity is 
generalised to quantal stochasticity, governed by a measure which satisfies a quantal analogue of the 
Kolmogorov sum rules \cite{Sorkin:1994dt}. For the purposes of this article, all that we need is the 
idea that the quantum dynamics for causal sets will be a generalisation of a stochastic process and so, as a ``warm up
exercise'',
it is natural to look first for physically motivated classical stochastic processes for dynamically generating causal sets. 

 A major development in this direction was the discovery by Rideout and Sorkin of a class of stochastic models, the Classical 
 Sequential Growth models, which satisfy certain physically natural conditions 
  \cite{Rideout:2000a}. Each model is a stochastic process in which a causal set grows 
 by the continual, spontaneous, random accretion of newly born elements. To describe the process requires the introduction of ``stages'' labelled by the  natural numbers.
 At stage $N$ of the process, there is already in existence a partial causal set of cardinality $N$ and a new causal set element is
 born. It chooses, according to the probability distribution that defines the process, which of the already existing causal set elements 
 to be to the future of, thus forming a new partial causal set of cardinality $N+1$.  This can be thought of as a 
 transition from one causal set to another with one more element, with a certain transition probability. 
 Then at stage $N+1$ another causal set element is born -- another transition occurs -- and so on. 
 
 Note that, at any fixed stage $N$, the given causal set may have been grown in different ways. For example, Figure 1 shows 
 two instances of the same 10 element causal set (the elements are the dots and the relations are indicated by the lines with the convention that  an upward-going sequence of lines from element $x$ to element $y$ indicates that $x$ precedes $y$). The 
labelling of each instance of the causal set indicates the stage at which each element is born and the two labellings are
different. However, to do justice to GR in which there is no global time coordinate, 
the labels on the causal set elements should have no physical significance -- in physics jargon, they should be pure gauge.
 That is, only the causal order indicated by the 
structure of relations between elements is physical. So, for example,  in labelling A the element 4 corresponds to the same physical entity as the element 8 in labelling B.

 This physical equivalence between the causal set labelled according to A and
and that labelled according to B implies that 
the transition probabilities of the Sequential Growth models should satisfy the condition that the probability of growing the 
causal set in the order indicated by labelling A must equal the  probability of growing the 
causal set in the order indicated by labelling B. This is the analogue in this discrete setting of 
 General Covariance in GR which implies that the actions for two spacetimes related by coordinate 
 transformations must be equal because they are physically equivalent. 

The Sequential Growth models satisfy this condition as well as a condition of 
\textit{Bell Causality} which is an expression -- appropriate in the classical case -- 
of the principle that the structure of the causal set in one region, $R$, should not 
affect the growth dynamics in a part of the causal set that is not related to $R$. Rideout and Sorkin find the most general 
form of the transition probabilities with these properties.

\begin{figure}[ht]
\begin{center}
\includegraphics[width=0.6\textwidth]{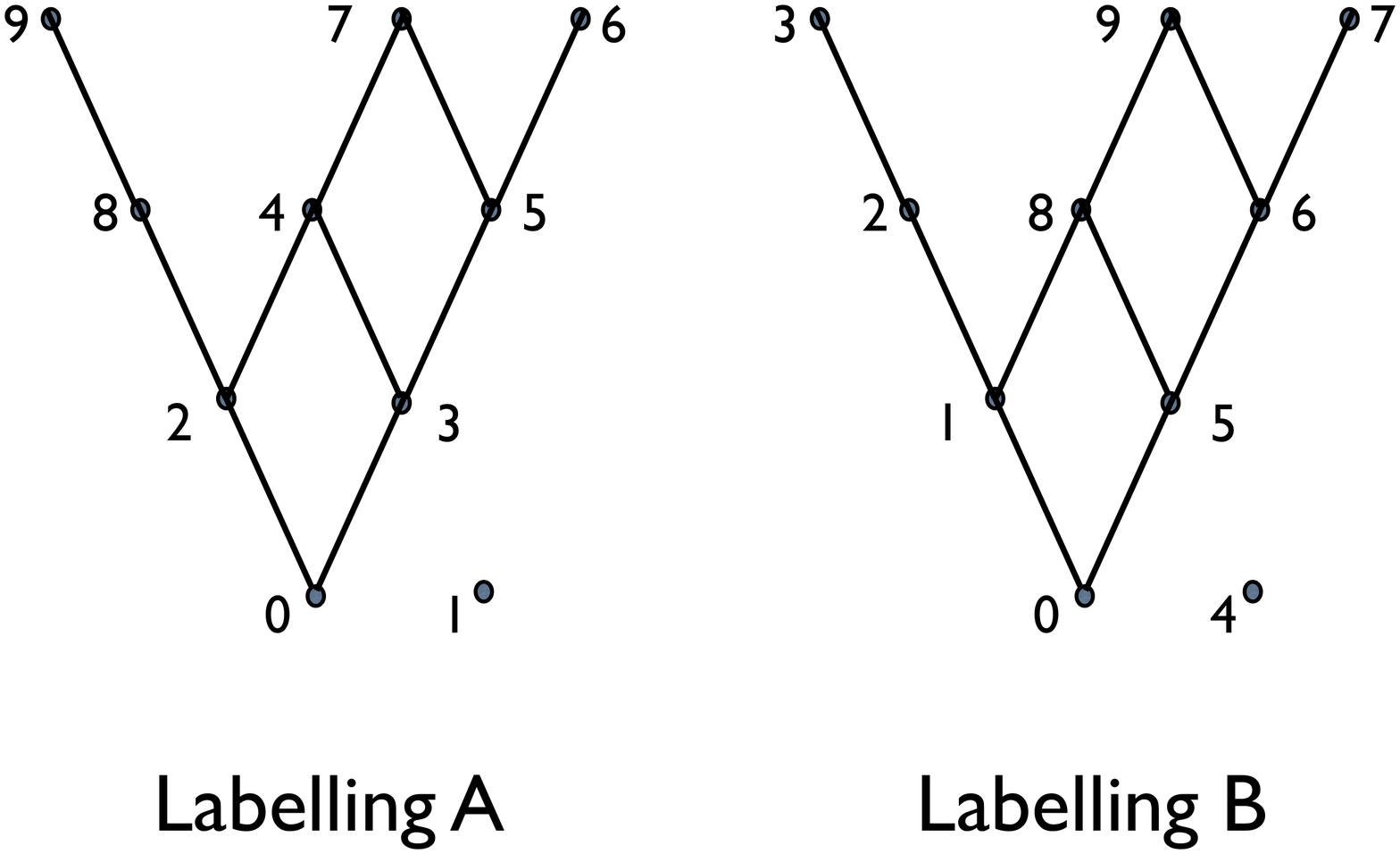}
\caption{Two labellings of the same partial causal set}
\label{growth}
\end{center}
\end{figure}

The Sequential Growth models have proved to be an important testing ground for ideas that will feed
 into the quantum generalisation. For example, the problem of identifying all the physically meaningful -- \textit{i.e.} label independent -- statements about a causal set has been solved for these models \cite{Brightwell:2002yu,Brightwell:2002vw}. For the
 question of interest in this article, however, the main significance of the models is as a proof of concept. A sequential growth model is a model of
 a physical world which \textit{becomes} in a manner compatible with the lack of a physical global  time. The \textit{physical} order in which elements of the causal set are born is the order they have in the 
 resulting causal set.
  For example in labelling A, the element labelled 2 is born before the element labelled 4 because 2 precedes
 4 in the causal set order. However,  there is no 
 physical order of the births of the elements labelled 2 and 3 because they are not ordered in the causal set order. 
  Sorkin  calls this partially ordered growth \textit{asynchronous becoming}. 

Sequential growth models bring a novel concept to physics: the partially ordered birth
process. This provides a possibility to locate the concept of the ``present'' in
the physics:  the Present is the \textit{Process}.  Each spacetime atom is born and it is its \textit{birth} rather than 
its \textit{being} that corresponds to the local, present moment. The local present moment is the 
birth of the atoms, not any collection of atoms themselves. 
 Being is a property of the past and once an atom has been born it is part of the past. 
Moreover, Sorkin points out \cite{Sorkin:2007qh}, our conscious experience is of a localised
present moment and we can correlate our experience with this birth of new spacetime atoms. 
Let me emphasise that conscious experience is not being treated here as different in kind to any other 
physical phenomenon. Every event, on this view,  has the same character, for example a tree falling in a forest
or a supernova exploding in a distant galaxy. The occurrence of the event is the partially ordered birth of the spacetime 
atoms (and, if necessary, accompanying matter degrees of freedom born with or on them) which comprise that 
event. The occurrence of an event is not the event itself. 

In making the conceptual 
distinction between an event and its occurrence,  it may be helpful to think of the description of the event as it is given in a Block 
Universe as corresponding to the event \textit{having happened}.  
A future event is conceived of in the Block as an event that \textit{will have happened}, 
not \textit{will happen}, a subtle but crucial distinction.
What is missing within a Block universe view is the \textit{occurrence} of the event: there is nothing 
in the Block to correlate with ``the event happens."  The birth process of sequential growth models
provides the missing concept, breathing life into the Block's static view of the physical world.  

A Blockhead can always play 
the card of ``running the process to infinity'' and conceiving of the model in a 
static way as a once-and-for-all random choice of one particular Block from a collection of possible
Blocks according to the probability distribution provided by the process. 
 My view on this manouevre is that it produces a \textit{different} physical theory 
-- some might want to call it a different interpretation of the theory -- in which there is nothing to 
correlate with the occurrence of events. This Block-ified theory \textit{correlates less well} 
with our sense-experience than the one in 
which the birth of new spacetime atoms is a physical process.   When there are 
different interpretations of a theory, it can presage scientific development in which one 
conception leads to progress that is obstructed by the other. 
We do not yet know if or how asynchronous becoming will be manifested in a  \textit{quantum} dynamics for causal set theory so 
it is too early to say if the lessons of the Rideout-Sorkin models can be taken as providing a solution to
 the problem of passage in physics. 
However, Sorkin anticipates that, in seeking the
quantum dynamics of causal sets,  championing Becoming over Being 
will be a fruitful heuristic. 

\vskip .5cm   \noindent\textbf{Final Thoughts}

Sorkin's suggestion that  the birth of spacetime atoms correlates with our conscious perception  challenges the view that quantum gravity effects can only show themselves as phenomena in 
regimes far beyond our immediate reach. The idea is that we may have access, through our intimate experience, to a physical phenomenon that is not present in General Relativity but \textit{is} part of a more complete theory of spacetime.  Lest this seem far-fetched, let me introduce  an example from history where an everyday human experience was pointing the way to a new understanding of the physical world, but the lack of a theoretical framework in which to situate and understand this experience meant that it remained (as far as I know) un-articulated until the beginning of the twentieth century. 

Every day in the centuries between Newton's discovery of the Law of Universal Gravitation and Einstein's discovery of General Relativity, every human being was making an observation with  no explanation within 
 the Newtonian theory but which correlates perfectly with GR. This 
 observational fact, was, throughout that time in history, ``hidden in plain sight.'' I invite you to make the observation yourself.
Sit down, close your eyes and spend a few minutes becoming aware of all the sensations on and within your
 body, with as little ``mental commentary'' as possible. One of the most
obvious sensations is the pressure of your chair upwards on your bottom. One thing that you do \textit{not} feel is a gravitational force (what we call ``weight'')  acting downwards on you, though the Newtonian theory tells us that there is such a force: there is a lack of correlation here between experience and Newtonian gravitational theory. 
In General Relativity, on the other hand, there is perfect correlation between experience and theory because in GR there 
\textit{is} no gravitational 
force acting downwards on you: we are always, wherever we are, weightless. Perhaps 
one reason GR is so immensely satisfying to learn 
is that it accords with our experience in this way.\footnote{I find 
it interesting that school students often 
make the mistake of forgetting to include the force of weight on bodies in mechanics problems: 
one might conjecture that they make this ``mistake''
 because of lack of experience of such a force, in contrast to reaction forces, tension, pressure \textit{etc.} which \textit{can} be felt.}  The phenomenon of lack of sense-experience of a downwards force of 
 weight occurs in everyday situations far removed from the physical regimes of strong curvature in which the full theory of GR reveals itself. Sorkin is suggesting that partial evidence for a theory of 
 quantum gravity may be similarly close to us, 
 although the full theory of quantum gravity is expected to 
 manifest itself only in extreme regimes. 
 
The experience that a gravitational force downwards is not
felt did not rule out the Newtonian physical world view; it did not even count as an observational fact because the theoretical framework in which it can be understood did not yet exist.\footnote{If there is only 
a force upwards on you from your chair and no weight acting down on you then you must be 
accelerating upwards, away from the earth. And so must someone sitting on a chair on the other side
of the planet. For everyone on the surface of the earth to be stationary 
and yet also accelerating away from the centre of the earth requires that the spacetime around the earth be curved.}
In the Newtonian world view, the feeling of the force upwards on you proves there must be 
a gravitational force of weight downwards on you, since you are at rest. The feeling of upward pressure is therefore interpreted as the appropriate experience of the force of weight
 downwards,  hence the terminology of ``weightlessness'' to label the experience of astronauts in space
 but not our experience on earth. 
 I am aware that this example is threatening to lead into a debate about whether sense-experiences can be 
completely theory-independent. My own opinion is that they  cannot; but what we \textit{can} 
decide is whether our sense-experiences find closer correlation with theory X or theory Y 
which is all that is necessary for our discussion.
Now, consider this fictitious conversation:

\begin{itemize}
\item[] \textit{17th Century Scientist}: There is a physical force of weight on you. Look at all the data, celestial mechanics, \textit{etc}. The Newtonian  theory of gravitation accounts for all that data.
\item[] \textit{17th Century Sceptic}: But I don't experience this gravitational force of weight whereas I can feel mechanical forces of comparable magnitude. Why?
\item[] \textit{Scientist}: The force of weight is physical. So your sense-experience of no force must be an illusion. Neurology, psychology, the way the mind and body work to produce sense-experience 
must be responsible for this illusion of lack of gravitational force of weight. 
\item[] \textit{Sceptic}: Maybe. But maybe this is telling us to look again at our theory, with the lack of gravitational force of weight as a heuristic. 
\end{itemize}

 The following is a parallel conversation that may, in the future, make similar sense:
 
\begin{itemize}
\item[] \textit{21st Century Blockhead}: There is no physical passage of time. Look at all the data, celestial mechanics, \textit{etc}. The theory of General Relativity with spacetime as a Block perfectly accounts for all that data.
\item[] \textit{21st century Sceptic}: But I don't experience a Block. I experience a sequence of moments. Why?
\item[] \textit{Blockhead}: The Block is physical, the passage of time is not physical. 
So your sense-experience of time passing must be an illusion. Neurology, psychology, the way the mind and body work to produce sense-experience must be responsible for this illusion of the passage of time. 
\item[] \textit{Sceptic}: Maybe. But maybe this is telling us to look again at our theory, with a physical
passage of time as a heuristic. \footnote{One could try to develop the analogy further. The  lack of a local gravitational force -- the 
equivalence principle -- 
is not the whole of GR, it is only one aspect of it. A great deal of further work needed to be done to 
arrive at GR, particularly on the precise form of the dynamical laws  governing the new spacetime substance.
The concept of the birth of spacetime atoms would be only one aspect of quantum gravity: the full theory including its quantum dynamical laws remains to be  discovered.} 

\end{itemize}

I will end by confessing a certain ambivalence towards my own claim that 
we may be able to comprehend the passage of time scientifically. 
Science requires communication of observational facts,  which communication
 requires or creates records of the facts. Mathematics is the
grand organising scheme used in physics to systematise the recorded data and to deduce new facts from
established ones using agreed upon rules of inference.
The scientific business of recording, organising and communicating 
seems to rob the sense-impression-of-time-passing of its essential qualities, that it is dynamic and fleeting. 

A. Einstein, in his Autobiographical Notes \cite{Schilpp:1949},  begins the statement of his
epistemological  credo thus: ``I see on the one side the totality of sense-experiences, and, on the other, the totality of the concepts and propositions which are \textit{laid down in books}'' (my emphasis). 
Einstein continues, ``A system [physical theory] has truth-content according to the certainty and completeness of its co-ordination-possibility to the totality of experience,'' which co-ordination is an intuitive act and ``not of a logical nature.'' 
 In struggling to describe the passage of time theoretically, we are seeking to correlate our 
 dynamic sense-experiences with something static, something laid down in 
 books. We may be coming up against a
fundamental inability of mathematics as we currently know it to co-ordinate with the physical world adequately. 
I agree with Sorkin that the sequential growth models come as close as one can 
imagine within current mathematics 
to realising a model of the universe
which reflects the dynamic character of our experience. Sorkin's metaphor of
\textit{ birth}  does the work of the intuitive, non-logical ``co-ordination'' between the 
sense-experience 
of new things happening and the mathematical, theoretical concept of a stochastic process.  
And yet it still perhaps misses 
something. Unless we are able to fashion mathematics to be more in tune with the dynamic 
quality of the physical world as we experience it,  we may have to continue to 
turn to music, dance and drama, 
rather than science, to express more fully our experience of the inexorable passage of time.

\section*{Acknowledgements} 
I am grateful to Jeremy Butterfield for a critical reading and for suggesting several improvements.
This work was supported in part by COST Action MP1006 Fundamental Problems in Quantum Physics. 
\begin{appendix}
\end{appendix}

\providecommand{\href}[2]{#2}\begingroup\raggedright\endgroup

\end{document}